\documentclass{JINST}
\let\ifpdf\relax
\usepackage{enumerate}

\title{ A Piggyback resistive Micromegas }

\author{ D. Atti\'e $^a$, A. Chaus $^a$, P.  Colas$^a$, E. Ferrer Ribas $^a$, J. Gal\'an $^a$, I.  Giomataris$^a$ \thanks{Corresponding author.}, F.J.~Iguaz $^a$, A. Gongadze $^a$, R. De Oliveira $^b$, T. Papaevangelou $^a$, A. Peyaud $^a$\\
\llap{$^a$}IRFU, CEA-Saclay,\\
  91191 Gif-sur-Yvette, France\\
\llap{$^b$}CERN,\\
  Geneva, Switzerland\\
  E-mail: \email{ioanis.giomataris@cern.ch}}

\abstract{ A novel read-out architecture has been developed for the Micromegas detector. The anode element is made of a resistive layer on a ceramic substrate. The detector part is entirely separated from the read-out element. Without significant loss, signals are transmitted by capacitive coupling to the read-out pads. The detector provides high gas gain, good energy resolution and the resistive layer assures spark protection to the electronics. This assembly could be combined with modern pixel array electronic ASICs. This readout organization is free on how the pixels are designed, arranged and connected. We present first results taken with a Medipix2 read-out chip.  }


\begin{document}

\section{Introduction}
One of the most prominent problems associated with gas-filled proportional chambers is the sparking induced by heavily ionizing particles producing large deposits. Amplified by the avalanche process they could reach a critical charge density, related to the Raether's limit \cite{bib1}, and could evolve into a discharge. As a result, a large fraction of the stored charge defining the amplification field is lost. The presence of discharges limits the high rate operation of the detector due to the required power supply recovery time, reduces the mean life of the detector and risks the damage of the readout electronics due to the high currents reached by these spark processes.

\vspace{0.1cm}
In Micromegas (MM) detector \cite{bib2,bib3} discharges are not destructive but they must be
reduced in order to obtain a safe operation.
\vspace{0.1cm}

To avoid any damage of the electronics, most of the detectors are using an additional protecting circuit, which interfaces the readout strips or pads with the front-end electronics. Recently new developments are undertaken to further improve the spark protection to operate at higher rates. A first promising approach is employing a resistive foil on top of the anode plane \cite{bib4}. When a streamer develops the increasing number of charges is deposited at the high resistivity foil. The anode potential generated by these charges reduces the amplification field, quenching the streamer resulting from field loss. A required condition is that the anode potential must last for a period of time long enough such that the charges present in the streamer have been completely evacuated, which introduces the need for high resistive values. The resistive foil limits the charge transfer, protecting the electronics, and avoids the total discharge of the mesh reducing the dead time of the detector.
\vspace{0.1cm}

Protection against discharges becomes mandatory in the case of GridPix detectors, where the Micromegas is placed directly on a pixelized silicon chip. The damage can be due to the plasma that locally melts or evaporates the chip material, or by a breakdown of electronic circuitry, due to too high potentials or charges. Even a local damage will not affect only a single channel, as it is the case in conventional read-out, but the whole chip will be damaged. Moreover, since the Micromegas is integrated in the chip, the whole structure must be replaced. In order to protect the chip from discharges a high-resistive layer of 5 $\mu$m up to 25 $\mu$m of amorphous silicon is deposited on the chip \cite{bib5,bib6}.
\vspace{0.1cm}

For large Micromegas detectors, a novel protection scheme has been developed by a group working on hard radiation detectors at CERN \cite{bib7,bib8}. According to this scheme, we are using resistive strips above the readout electrode to overcome the spark problem. This is a quite efficient and robust solution for fabricating very-large area detectors.
\vspace{0.1cm}

We present a new approach called "Piggyback" resistive Micromegas where a thin resistive layer is deposited on an adequate insulator. This element is used as substrate for the Micromegas which could be fabricated for instance by using the "bulk" technology~\cite{bib9}.

\section{The new technology description}

\begin{figure}[htbp]
\begin{center}
\includegraphics[width=7.5cm]{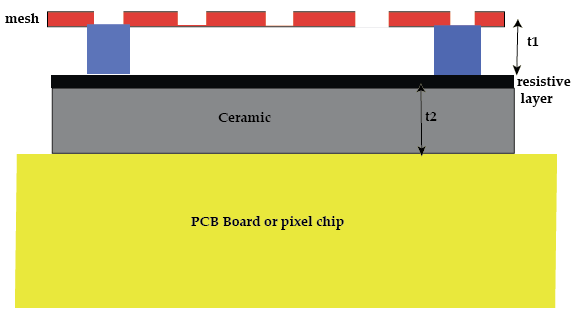}
\caption{From top to bottom, the MM structure with thickness t$_1$ (mesh in red, pillars in blue), the resistive layer (in black), the ceramic with thickness t$_2$ and the PCB Board or pixel chip.}
\label{fig1}
\end{center}
\end{figure}

The new idea proposed consists in separating the detector part from the read-out part as shown in figure~\ref{fig1}; the signal is transmitted by capacitive coupling to the read-out pads. The technology used here is inspired from a similar development on a Parallel Plate Avalanche Counter described in reference~\cite{bib10}.

In the proposed technology the amplification gap (t$_1$) is about 100 $\mu$m and it is achieved between a woven stainless steel mesh connected to the high voltage and a resistive anode. In some special configurations the anode element could be also connected to high voltage. By adding a drift electrode, a MM detector operates as usually in the proportional avalanche mode inducing signals on the resistive anode plane.
Various elements of the structure are optimized in such a way that the electronic signal is not lost through the resistive layer but is propagated to a separate plane, carrying read-out pads or strips, by capacitive coupling. A schematic view of the new structure is shown in figure~\ref{fig1} where we can see the various elements: the Micromegas structure with the pillars; the resistive layer; the ceramic substrate and the readout board. The read-out plane can be a simple PCB board or a pixel chip such as the Medipix2 \cite{bib6} or any similar integrated chip. The thickness of the resistive layer is of the order of 10 $\mu$m.

\vspace{0.1cm}
In order to optimize the induced signal by capacitive coupling the thickness of the insulator (t$_2$) must be kept small to satisfy the relation

\begin{equation}
t_2 \ll t_1 \frac{\epsilon_2}{\epsilon_1}
\label{permRel}
\end{equation}

\noindent where $\epsilon_1$ is the dielectric constant of medium 1 (gas) and $\epsilon_2$ is the dielectric constant of medium 2 (insulator). Because the insulator plays at the same time the role of vessel of the detector the thickness should be kept reasonable (several hundred $\mu$m). In order to satisfy \ref{permRel} the material should have a dielectric constant as high as possible. Good candidates are ceramic insulators having large dielectric constants ($\gg$10).  In our first prototype we have used a standard amplification gap of t$_1$=128\,$\mu$m, the ceramic insulator was alumina with t$_2$=300\,$\mu$m ($\epsilon_1$=10). The selection of the resistive layer is usually a challenge in gas detectors in order to find an appropriate resistivity and a good surface quality. The ruthenium oxide (RuO$_2$) has been chosen. This latter is extensively used for coating ceramics at high temperature for the preparation of resistors or integrated circuits.

\vspace{0.1cm}

The RuO$_2$ based thick film was prepared by a standard screen printing technique from a commercial paste having a resistivity of 100 M$\Omega/\square$. This is a cost effective industrial process that permits to get a robust and stable layer with a variety of values for the resistivity: 1 M$\Omega$ to hundreds of G$\Omega$. The combination of the ceramic insulator, having an excellent surface quality and the robustness of the resistive layer, provides a solid solution for our structure.

\section{Experimental set-up and results}

\begin{figure}[htbp]
\begin{center}
\includegraphics[width=7.5cm]{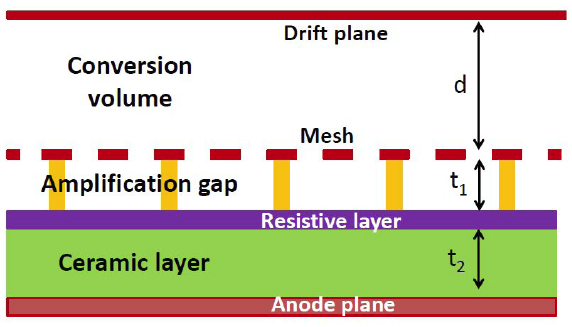}
\caption{Schema of the Piggyback detectors}
\label{fig2}
\end{center}
\end{figure}

Three small detectors, with the schematic shown in figure~\ref{fig2}, were fabricated for these tests. The ceramic substrate was made out of standard alumina of 300 $\mu$m-thickness (t$_2$). A 20 $\mu$m thick RuO$_2$ based film with a resistivity of 100 M$\Omega/\square$ was processed above it. The backside of the ceramic layer was covered by a copper layer, which acted as the anode plane, and a bulk Micromegas structure \cite{bib9} was built on the resistive layer, with an amplification gap of 128\,$\mu$m (t$_1$). The whole structure was situated inside a gas chamber, specifically designed for these tests, and a mesh frame was used as the drift plane, in order to define a 10 mm thick conversion volume.

From the three detectors built, only the number 1 and 3 worked during the tests. They were illuminated by a $^{55}$Fe source keeping a constant gas flow. The mesh voltage was varied from 300 to 450\,V, the drift from 400 to 1200\,V and the anode was connected to ground. Mesh and drift voltages were independently powered by a CAEN N471A module. The avalanche in the Micromegas structure induces a negative signal in the mesh and a positive one in the anode. Both signals were respectively monitored by two CANBERRA preamplifiers to evaluate their efficiency. The preamplifier outputs were fed into two CANBERRA 2006 amplifiers and subsequently into a multichannel analyzer AMPTEK MCA-8000A for spectra acquisition. Each spectrum was fitted by two Gaussian functions, corresponding to the K$_\alpha$ (5.9 keV) and K$_\beta$ (6.4 keV) lines of the $^{55}$Fe source; the mean position and width of the main peak were obtained for each fit. 

\begin{figure}[htbp]
\begin{center}
\includegraphics[width=7.5cm]{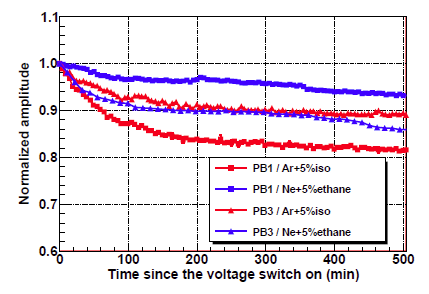}
\caption{Gain evolution along time since the voltages have been switched on for the Piggyback detectors 1 (squared line) and 3 (triangled line), respectively tested in Ar+5\%iC$_4$H$_{10}$ (blue) and in Ne+5\%C$_2$H$_6$ (red).}
\label{fig3}
\end{center}
\end{figure}

\vspace{0.1cm}
Two types of gas mixtures were used in these tests: Argon + 5\% Isobutane (iC$_4$H$_{10}$) and Neon + 5\% Ethane (C$_2$H$_6$). The voltage was then switched on and the evolution of the detector's gain was studied generating an energy spectrum integrated for 1 minute every 5 minutes during several hours. As shown in figure~\ref{fig3}, the gain of this type of detectors decreases during the first 100 minutes, until a stable operation point is reached. The gain may later change due to pressure and temperature variations, as the amplification gap of 128 $\mu$m is not the optimum for the gas mixtures used \cite{bib2}.

\begin{figure}[htbp]
\begin{center}
\includegraphics[width=15cm]{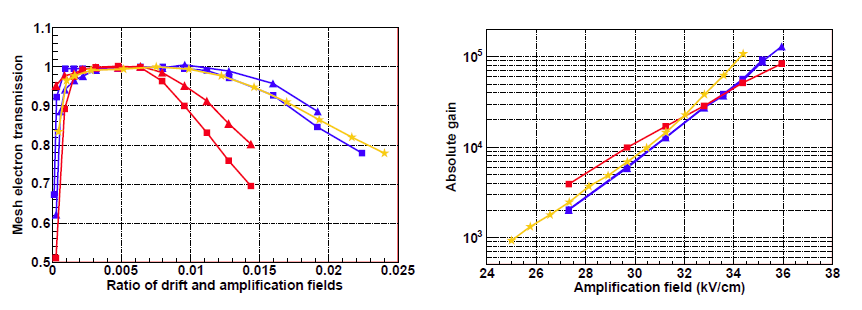}
\caption{Dependence of the electron mesh transmission with the ratio of the drift and amplification fields (left) and gain curves (right) for the Piggyback detectors 1 (squared line) and 3 (triangled line), respectively tested in Ar+5\%iC$_4$H$_{10}$ (blue) and in Ne+5\%C$_2$H$_6$(red). The curves of a 128 $\mu$m-thickness gap bulk detector (orange stars) in Ar+5\%iC$_4$H$_{10}$ extracted from \cite{bib7}, have been added as a comparison.}
\label{fig4}
\end{center}
\end{figure}

Once the gain has stabilized, the drift voltage was varied for a fixed mesh voltage to obtain the electron transmission curve, shown in figure~\ref{fig4} (left). As a typical Micromegas detector, these curves show a "plateau" of maximum electron transmission for a range of low drift fields. For more intense fields, the mesh stops being transparent for the primary electrons generated in the conversion volume and both the gain and the energy resolution degrade. The ratio of drift and amplification fields was then fixed to an optimum and the mesh voltage was varied. The dependence of the peak position with the mesh voltage generates the gain curve, shown in figure~\ref{fig4} (right). Both detectors showed slightly lower gains than a standard 128 $\mu$m-thickness bulk detector for the same mesh voltages but they both reached values up to 10$^5$ before the spark limit.

\begin{figure}[htbp]
\begin{center}
\includegraphics[width=15cm]{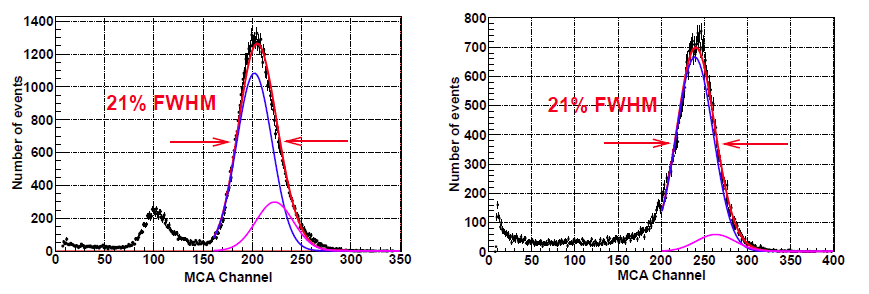}
\caption{Energy spectra generated by the MCA irradiating the Piggyback detector 1 by a $^{55}$Fe source respectively in Ar+5\%iC$_4$H$_{10}$ (left) and Ne+5\%C$_2$H$_6$ (right).  The main peak has been fitted to two Gaussian functions (blue and magenta lines), corresponding to the K$_\alpha$}
\label{fig5}
\end{center}
\end{figure}

\newpage
The energy resolution of both detectors was 21\% (FWHM) at 5.9 keV in Ar+5\%iC$_4$H$_{10}$ and in Ne+5\%C$_2$H$_6$, as shown in figure~\ref{fig5}. These values are slightly worse than for bulk detectors, which can reach values 16\% (FWHM) at 5.9 keV in argon-isobutane mixtures, but discharges are highly suppressed.

\begin{figure}[htbp]
\begin{center}
\includegraphics[width=15cm]{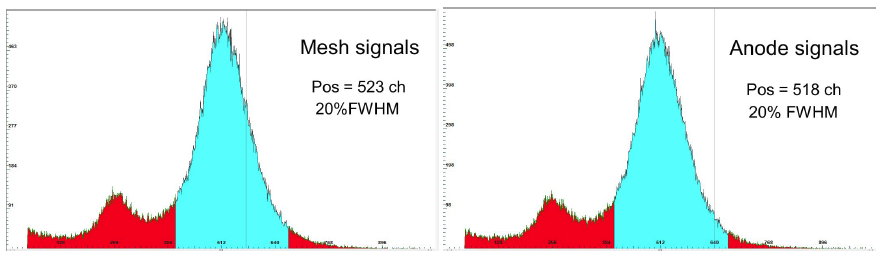}
\caption{Energy spectra generated by mesh (left) and anode's pulses (right) when the Piggyback detector 1 was irradiated by a $^{55}$Fe source in Ar+5\%iC$_4$H$_{10}$.}
\label{fig6}
\end{center}
\end{figure}

Finally, the signal induced both in the mesh and the anode plane was compared to evaluate possible losses at the anode plane by the thick ceramic layer. As shown in figure~\ref{fig6}, the signal was entirely propagated to the induction plane and the energy spectrum of both the mesh and the anode planes showed similar values, with an efficiency better than 90\%.

\begin{figure}[htbp]
\begin{center}
\includegraphics[width=7.5cm]{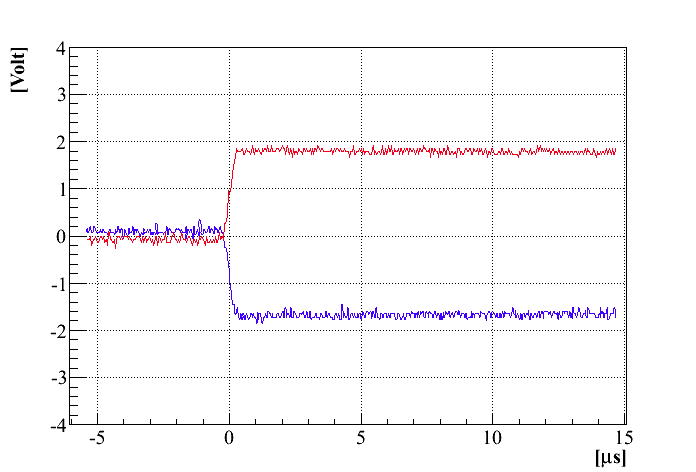}
\includegraphics[width=7.5cm]{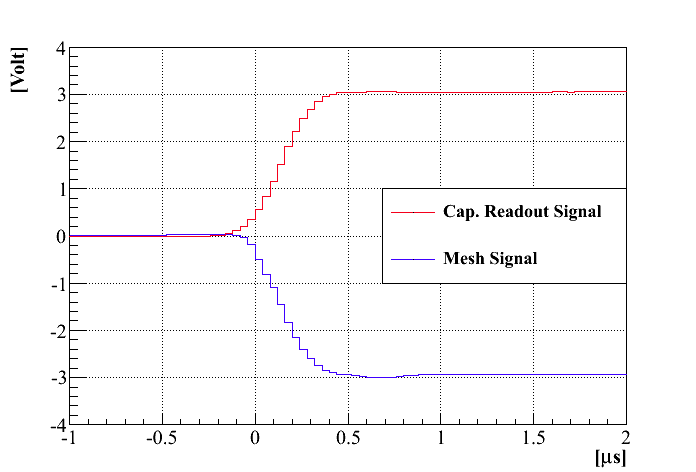}
\caption{Signals from fission fragments as they are recorded by the mesh and the anode.  On the left a single event is drawn, while on right is shown the accumulation of 5000 events.}
\label{fig7}
\end{center}
\end{figure}

\vspace{0.1cm}
In order to minimize the -possibly- different behavior of the electronics chain to positive and negative signals, we repeated the measurement using a $^{252}$Cf source and two Ortec 142B preamplifiers registering the mesh and anode signals. The signals were registered simultaneously while the measurement was repeated after interchanging the preamplifiers to exclude systematics. The fission fragments deposit enough energy to create very big signals. Their range in the argon gas is of the order of 1 cm, so the rise time is expected to be of the order of 200-300 ns. A typical signal recorder simultaneously by the mesh and by the anode plane is shown on figure~\ref{fig7}. At the same figure is also shown the accumulated signals from 5000 events.

\begin{figure}[htbp]
\begin{center}
\includegraphics[width=7.5cm]{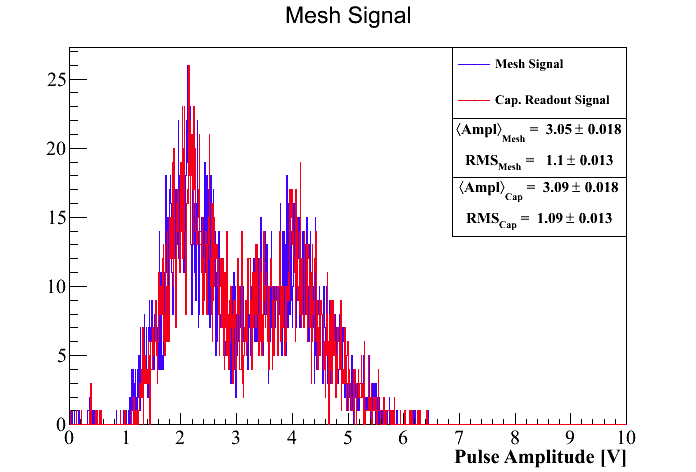}
\includegraphics[width=7.5cm]{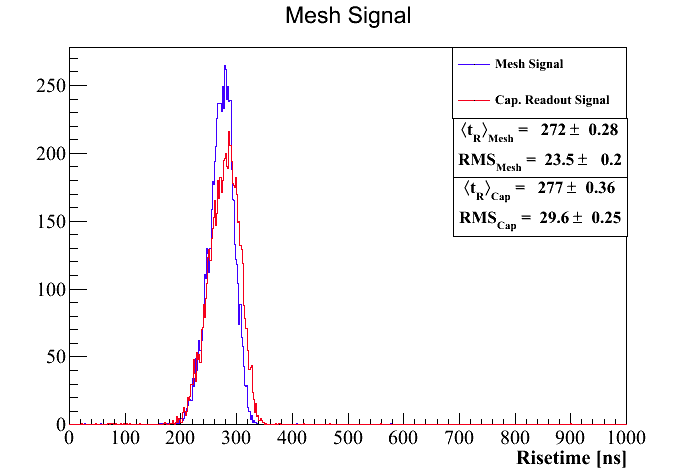}
\caption{Pulse amplitude (left) and risetime distribution from $^{252}$Cf fission fragment signals, as they were registered simultaneously from the mesh and the anode through capacitive coupling.  No significant loss is observed for both variables.}
\label{fig8}
\end{center}
\end{figure}

The amplitude and rise time distributions of the registered pulses are shown in figure~\ref{fig8}. The response of the preamplifiers to negative and to positive signals was measured with the help of a pulse generator and was found to be the same within 5\% (small loss for negative pulses). Given that fact and the results shown in figure~\ref{fig8}, we verify that the signal loss due to the capacitance of the ceramic is less than 10\%. The rise time of the pulses is also the same for both polarities within few percent, and does not appear to be affected by the resistive film, being within the expected range of 200-300 ns.

\newpage
\section{Rate capability}
Additional tests were performed using an X-ray generator which allowed the illumination of the Piggyback structure at several rates. The structure was placed inside a chamber with a drift gap of 0.5\,cm. The copper peak (8\,keV) from the X-ray generator cathode was used during these measurements as a gain reference. Given that the minimum generator intensity was already high a foil of about 100\,$\mu$m Cu and some additional aluminum was used to attenuate the overall flux.

\begin{figure}[htbp]
\begin{center}
\includegraphics[width=15cm]{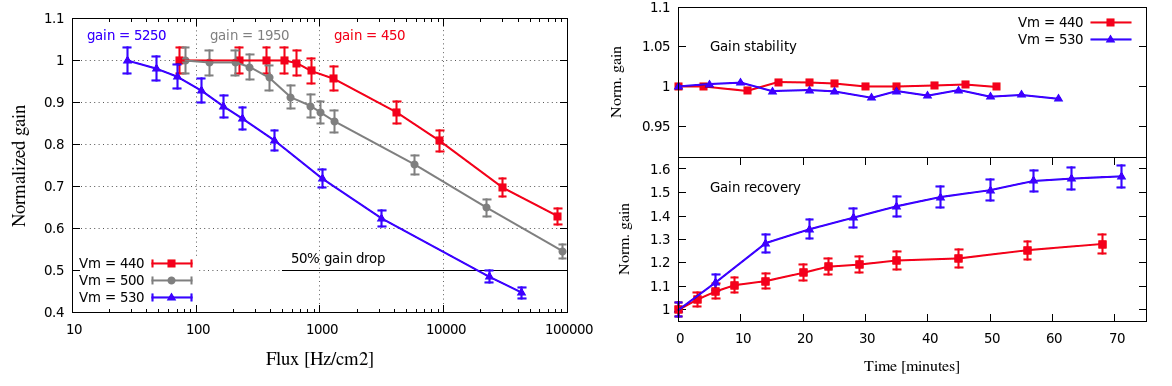}
\caption{Relative gain drop as a function of the X-ray flux and amplification gain (left), gain stability at the plateau region (right top) and gain recovery after high flux exposure to low flux transition (right bottom).}
\label{fig9}
\end{center}
\end{figure}

\vspace{0.1cm}
Gain measurements under the same conditions were done for different amplification gain values corresponding to different mesh voltages (440V, 500V and 530V) for X-ray fluxes up to about 100 kHz/cm$^2$, as observed in figure~\ref{fig9}.

\vspace{0.1cm}
These measurements show a "plateau" region dependent on the mesh voltage applied and that extends to higher fluxes for lower values of the amplification field. The gain stability was measured at low rates just after the system was started and the Piggyback illuminated with X-rays, showing gain stability better than 2\%.

For higher rates, the gain drops as a function of the flux and the time required to achieve a stable gain is in the order of several minutes. The recovery time from the highest rates applied to the "plateau" region is shown also in figure~\ref{fig9}.

\vspace{0.1cm}

It must be noticed that the gain dependency with the rate is not an intrinsic feature of the Piggyback structure and will also depend on the resistivity and effective ceramic capacitance chosen, allowing tuning up the values depending on the application, i.e.  lower values of the resistivity would lead to an extended "plateau" for higher rates and enhanced gain stability at higher rates.

\section{Simulation of charge diffusion effects over a resistive plate}

Charge diffusion over the resistive top plane could be one of the main causes of the observed behavior of the Piggyback. In this section we show that the accumulation of charge over time due to slow charge diffusion in the illuminated region could be affecting the measured gain and be directly related to the stabilization time required. The charge diffusion relation for a given charge $Q$, which follows a Gaussian distribution of width $w$, was provided in~\cite{bib17}

$$
\rho(r,t) = \frac{Q}{2\pi ( 2ht + w^2 )} \mbox{exp}\left[ -\frac{r^2}{2 (2ht + w^2)} \right] \,\,\,\,\,\,\,\,\,\,\,\,\,\,\,\,\,\,\,\,\,\,\,\,\,\,\,\,\,\,\,\,\,\, h=1/RC
$$

\vspace{0.2cm}
\noindent where the parameter $h$ it is related to the surface resistivity $R$ and surface capacitance $C$. The accumulated charge will be translated into a surface potential given by the capacitance of the ceramic and thus, in a reduction of the amplification field. In order to study such reduction a narrow region around the Gaussian distributed charge has been chosen. The evolution on time of the total charge inside this region, defined by the radius $R_o$, it is integrated from the previous expression and is described by the following relation

$$
\rho(R_o,t) = Q \left[ 1 - \mbox{exp}\left( -\frac{R_o^2}{2 (2ht + w^2)} \right) \right].
$$

\vspace{0.2cm}
We can extrapolate this relation to the case of a continuous current flow by assuming this relation is valid for a differential element of charge. Integrating these differential contributions over time will allow us to calculate the accumulated charge at any time,

$$
Q(R_o,t_o) = \int_0^{t_o} \frac{d\rho(R_o,t)}{dt} dt = \int_0^{t_o} \left[ 1 - \mbox{exp}\left( -\frac{R_o^2}{2 (2h(t_o - t) + w^2)} \right) \right] \left( \frac{dq(V_a)}{dt} \right) dt
$$

\vspace{0.2cm}
\noindent where we should take into account that these differential charge contributions will be related to the local amplification field, modified by the anode potential $V_a$ at the resistive plane at each time interval.

\begin{figure}[htbp]
\begin{center}
\includegraphics[width=8.5cm]{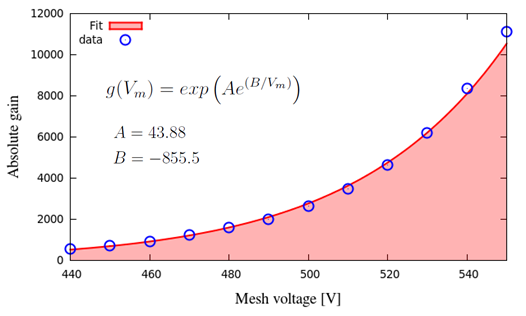}
\caption{Absolute gain measured (blue points) versus applied mesh voltage in Ar+10\% CO2, and the fitting result (red-filled curve).}
\label{fig10}
\end{center}
\end{figure}

\vspace{0.1cm}
The integral will be calculated by adding the charge contributions at each time interval n, and describing the gain in terms of the relative mesh versus anode potential,

$$
\delta q_n = g(V_a(n\delta t))N_e q_e r \,\,\,\,\,\,\,\,\,\,\,\,\,\,\,\,\, V_a(n\delta t) = q_n/\pi R_o^2 C
$$

\noindent where $N_e$ is the number of electrons produced in each interaction, $r$ is the interaction rate, $g$ is the gain as a function of the anode potential $V_a$. The gain as a function of the mesh potential is obtained experimentally by illuminating the Piggyback at low rates (see figure~\ref{fig10}), expression that will be used afterwards with the corrected potential given by V$_a$.

\begin{figure}[htbp]
\begin{center}
\includegraphics[width=15cm]{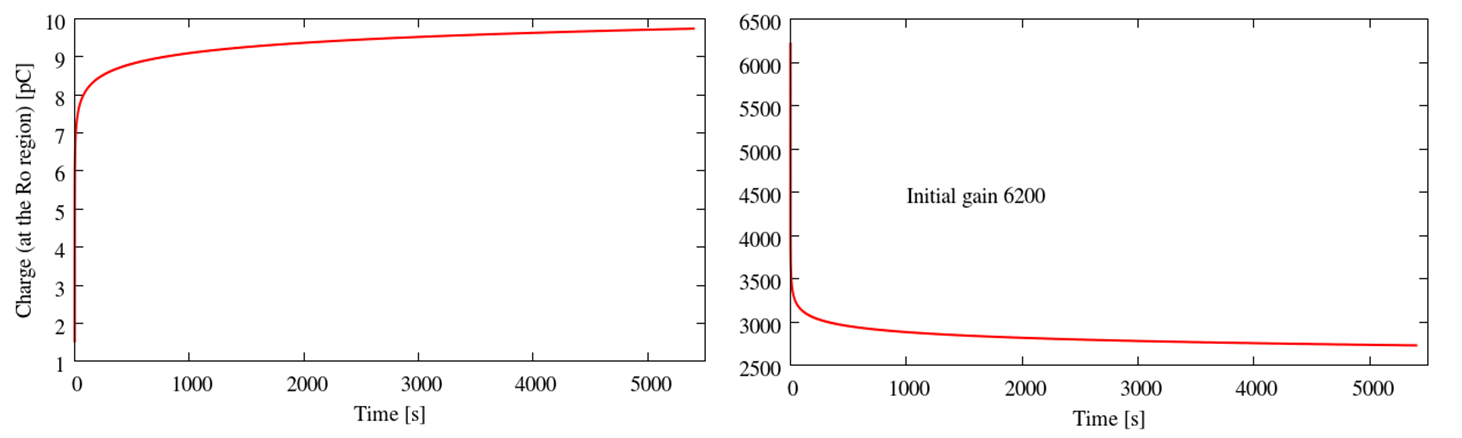}
\caption{Charge accumulated at the exposed region inside R$_o$, and the gain drop due to the voltage anode potential generated.}
\label{fig11}
\end{center}
\end{figure}

The charge is integrated considering the effect the accumulated charge will have in the gain at each time step. The simulation of each time step requires to know previously all the charges at each past time interval in order to apply the diffusion relation to them, this makes the computation of each time interval more expensive and the time required has a non-linear growth. The simulation of the first half an hour with reasonable accuracy can still be performed within a few hours. Figure~\ref{fig11} shows the charge evolution and the gain drop induced by a rate of 100 kHz/cm$^2$, for a resistivity of 100 M$\Omega/\square$  and 1pF/mm$^2$. As to compare with the experimental measurements we considered the number of electrons produced by an 8\,keV event, N$_e$ = 300 in an Ar+10\%CO$_2$ mixture.

This simulation shows a fast gain drop of about 50\% during the first seconds, and a second order drop of about 15\% the next one hour and a half. This behavior is compatible with that observed on figure~\ref{fig3}, considering that the first (fast) gain drop could not be observed in the measurement.

\begin{figure}[htbp]
\begin{center}
\includegraphics[width=15cm]{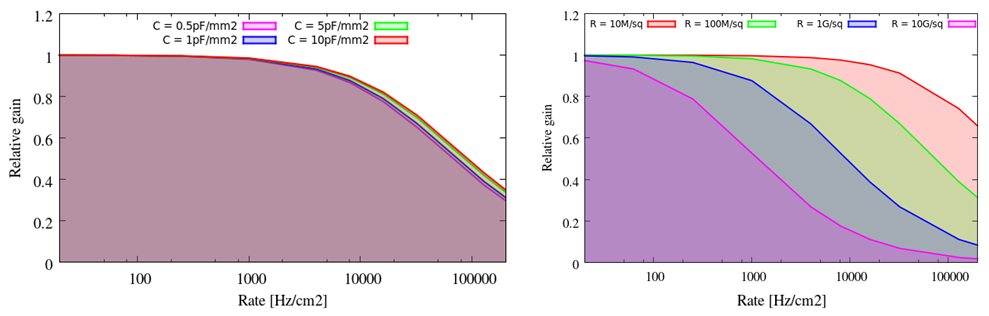}
\caption{On the left, relative gain versus the particle flux  for different resistivity values at 1pF/mm$^2$. On the right, relative gain versus flux for different capacitances at a resistivity of 100M$\Omega/\square$.}
\label{fig12}
\end{center}
\end{figure}

\vspace{0.1cm}
A set of simulations was launched at different scanning rates for different RC values in order to observe the effect on the gain drop observed. Figure~\ref{fig12} shows the final gain value obtained after an exposure of 5400 seconds, several simulations were performed at different rates in order to show the result as a function of the interaction flux. Different resistivity and capacitive values are tested from the range of 10M$\Omega/\square$  to 10\,G$\Omega/\square$, and from 0.5\,pF/mm$^2$ to 10\,pF/mm$^2$, showing the dependency with the applied rate for each case.

As lowest is the resistivity of the material, the longest is the flat gain region, and higher rates could be achieved without expecting an evolution in the gain, which of course, will also be faster for lower values of RC. The gain strongest dependency is with the resistivity of the foil, showing not strong dependency with the different capacitances simulated. However, the capacitance will have an important role in the charge diffusion time and thus in the stabilization time. 

\begin{figure}[htbp]
\begin{center}
\includegraphics[width=7.5cm]{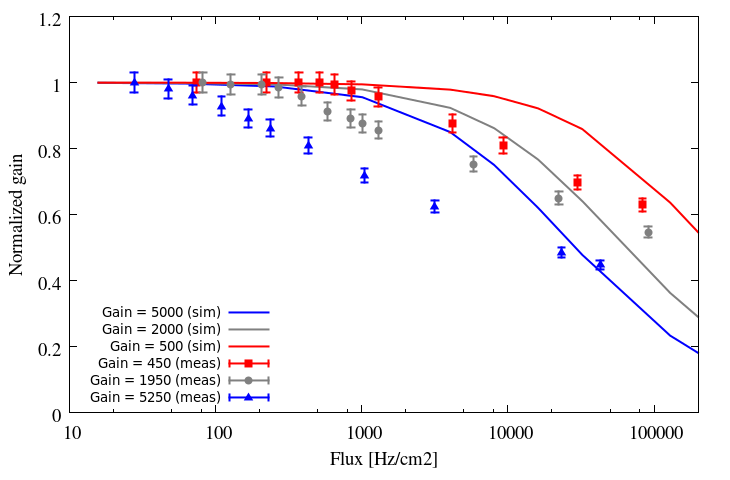}
\caption{Comparison of the experimental gain and the simulated gain drop dependency with the applied rate. The value of the resistivity used in the simulation was 500M$\Omega/\square$.}
\label{fig13}
\end{center}
\end{figure}

\vspace{0.1cm}
Finally, the measurements done with the X-ray generator allows to monitor the gain with the position of the 8\,keV peak. In figure~\ref{fig13} the charge diffusion observed on the gain could explain the contribution to the observed effect on the gain drop.

The agreement between simulation and measurement could be biased by other effects such as grounding configuration, real resistivity value and homogeneity, and systematics induced in the experimental measurements which are constrained by the fact that the measurements should be made in a relative short period of time, but they also require a reasonable stabilization time.
 
\section{CMOS chip read-out}
Medipix2 \cite{bib11} and Timepix \cite{bib12} are CMOS chips which have been used as pixelated readout for Micro-Pattern Gaseous Detectors (MPGDs)~\cite{bib6,bib13}. Both chips have the same size of 14$\times$16\,mm$^2$. They consist of a matrix of 256$\times$256 identical square pixels of 55\,$\mu$m side each. A regular topic associated to MPGDs is electronics protection to sparks. For this reason, layers of high resistivity material (amorphous silicon or silicon-rich nitride) have been deposited on CMOS pixel for discharge protection~\cite{bib14}. The Piggyback concept provides a spark protection by completely decoupling the detector from the electronics.

\vspace{0.1cm}
A 30$\times$20\,mm$^2$ bulk on a ceramic has been built to be used with Medipix2/Timepix chips. The amplification gap and the drift height were respectively 128\,$\mu$m and 10\,mm. The ceramic was put on top a Medipix2 chip without gluing. The first image was recorded (see figure~\ref{fig10}) using a $^{55}$Fe radioactive source. The mesh voltage was 430\,V corresponding to a gain of about 105 in standard bulk in a gas mixture of argon and 5\% isobutane. The detector was operating under such high gain for a whole day verifying the effectiveness of the spark protection scheme. The first results are extremely encouraging, demonstrating the proof of principle of this novel read-out architecture for the Micromegas detector.

\begin{figure}[htbp]
\begin{center}
\includegraphics[width=7.5cm]{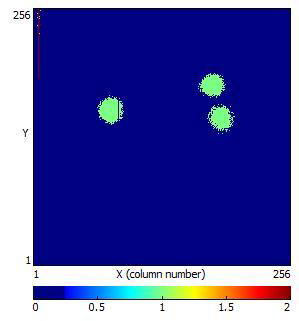}
\caption{Three events detected and recorded by the Medipix2 chip.}
\label{fig14}
\end{center}
\end{figure}

\section{Summary and outlook}

We have fabricated and tested a new Micromegas detector with resistive anode deposited by an industrial process on a thin ceramic substrate. The first results are promising; the detector is stable, spark protected, provides a high gas gain and good energy resolution.

\vspace{0.1cm}
The new structure provides several advantages and opens new windows of opportunity in particle detectors. In particular:

\begin{enumerate}[(i)]
\item The detector is completely dissociated from the read-out plane. This gives complete freedom to the user to choose the readout electronics card as a function of its needs; i.e. the card is easily adapted to the detector structure. As well it can be removed or replaced without switching off the detector

\item In the case of integrated pixel chips this structure should solve the technical problem related to the implementation of many chips, without creating dead
space, since no additional support of interface structure is needed.

\item In the case of ceramic insulator there is an additional advantage: ceramics can provide large values of the dielectric constant which can reach values up to several thousand. This opens the possibility to select, when it is necessary, a thick insulator without violating the relation~\ref{permRel}. With such large thicknesses the induced signal on the read-out strips could be spread-out to large values permitting the use of large size pads without degradation of the spatial resolution: the good spatial resolution is maintained by "pad charge sharing"~\cite{bib15,bib16}. This is a big advantage compared to conventional Micromegas which needs narrow high-density anode readout element to achieve good spatial resolution.

\item The proposed scheme provides a full spark protection. No additional protection structure is needed for the read-out electronic chip.

\item It is very easy to apply positive voltages on the micromesh and the anode and keep the cathode grounded, with no need of high voltage decoupling circuit for the anode signals. This can be important for the case that the cathode is also the entrance window.

\item The detector materials, ruthenium oxide and ceramics, exhibit excellent outgassing properties. This is appropriate for high quality vacuum and it opens the way to seal the detector vessel. 
\end{enumerate}

Future developments are under way to improve smoothness and build larger surfaces. For small surfaces we would like to decrease as much as possible the thickness of the ceramic to improve the efficiency in the case of small pad readout chips like the TimePix.

\vspace{0.1cm}
For large-size detectors we could use thick ceramic substrate. This will produce large spread of the signals and could allow the usage of large read-out pads for applications where cost savings on electronics is required. Very large detector areas could be built using several ceramic structures mounted into a mosaic. To obtain a good flatness, tilts of the surfaces must be minimized and dead area can be reduced to a negligible level by precisely adjusting the ceramic substrates and by fixing them on top of the read-out strip or pixel plane.  Interconnection of the various resistive films to the ground should be achieved through an appropriate technology that has to be developed. The Bulk process for fabricating the Micromegas gap could be applied over the global read-out area (mosaic).


\end{document}